\documentstyle[aps,preprint]{revtex}

\begin{document}
\title{Decay rate and renormalized frequency shift of a quantum wire Wannier
exciton in a planar microcavity}
\author{Yueh-Nan Chen and Der-San Chuu\thanks{%
corresponding author: e-mail: dschuu@cc.nctu.edu.tw; Fax: 886-3-5725230;
Tel: 886-3-5712121-56105 }}
\address{Department of Electrophysics, National Chiao Tung University,\\
Hsinchu 30050, Taiwan}
\author{T. Brandes}
\address{Department of Physics, UMIST, P.O. Box 88, Manchester, M60 1QD, U.K.}
\author{B. Kramer}
\address{University of Hamburg, 1. Institut. f\"{u}r Theoretische Physik,\\
Jungiusstrasse 9, D-20355 Hamburg, Germany}
\date{\today}
\maketitle

\begin{abstract}
The superradiant decay rate and frequency shift of a Wannier exciton in a
one-dimensional quantum wire are studied. It is shown that the dark mode
exciton can be examined experimentally when the quantum wire is embedded in
a planar microcavity. It is also found that the decay rate is greatly
enhanced as the cavity length $L_{c}$ is equal to the multiple wavelength of
the emitted photon. Similar to its decay rate counterpart, the frequency
shift also shows discontinuities at resonant modes.
\end{abstract}

\newpage

\bigskip

Historically, the idea of superradiance was introduced by Dicke\cite{1}.
Later, the coherent radiation phenomena for the atomic system was
intensively investigated\cite{2,3,4,5,6}. One of the limiting cases of
superradiance is the exciton-polariton state in solid state physics. But as
it was well known in a 3-D bulk crystal\cite{7},{\it \ }the excitons will
couple with photons to form polaritons--the eigenstate of the combined
system consisting of the crystal and the radiation field which does not
decay radiatively. If one considers a linear chain or a thin film, the
exciton can undergo radiative decay as a result of the broken crystal
symmetry. The decay rate of the exciton is enhanced by a factor of $\lambda
/d$ in a linear chain\cite{8} and ($\lambda /d)^{2}$ for 2D
exciton-polariton \cite{9,10}, where $\lambda $ is the wave length of
emitted photon and $d$ is the lattice constant of the linear chain or the
thin film.

First observation of superradiant short lifetimes of excitons has been
performed by Ya. Aaviksoo {\em et al}.\cite{11} on surface states of the
anthracene crystal. Later, B. Deveaud {\em et al}.\cite{12} measured the
radiative lifetimes of free excitons in GaAs quantum wells and observed the
enhanced radiative recombination of the excitons. Hanamura\cite{13}
investigated theoretically the radiative decay rate of quantum dot and
quantum well excitons. The results obtained by Hanamura are in agreement
with that of Lee and Liu's\cite{10} prediction for thin films. Knoester\cite
{14} obtained the dispersion relation of Frenkel excitons of quantum slab.
An oscillating dependence of the radiative width of the excitonlike
polaritons with the lowest energy on the crystal thickness was found.
Recently, G. Bj\"{o}rk {\em et al}.\cite{15} examined the relationship
between atomic and excitonic superradiance in thin and thick slab
geometries. They demonstrated that superradiance can be treated by a unified
formalism for Frenkel excitons and Wannier excitons. In V. M. Agranovich 
{\em et al}.'s work\cite{16}, a detailed microscopic study of Frenkel
exciton-polariton in crystal slabs of arbitrary thickness was performed.

For lower dimensional systems, A. L. Ivanov and H. Haug\cite{17} predicted
the existence of an exciton crystal, which favors coherent emission in the
form of superradiance in quantum wires. Y. Manabe {\em et al}.\cite{18}
considered the superradiance of interacting Frenkel excitons in a linear
chain. Recently, with the advances of the modern fabrication technology, it
has become possible to fabricate the planar microcavities incorporating
quantum wires\cite{19}. Although some of the theoretical papers discussed
the exciton-polariton splitting of quantum wires embedded in a microcavity 
\cite{20}, the spontaneous emission of the exciton as a function of cavity
length has received no attention. In this paper, we will investigate the
radiative decay of the Wannier exciton in one-dimensional quantum wires
embedded in planar microcavities. It will be shown that some interesting
quantities may be measured by making use of the properties of the
microcavity.

For simplicity, let us first approximate the quantum wire as a linear chain
with lattice spacing $d$ in a free space. As it was well known, the
Sommerfeld factor is smaller than unity in a one-dimensional system\cite{21}%
. The strong Coulomb interaction moves the oscillator strength out of the
continuum states into the exciton resonance. Practically the entire
oscillator strength is accumulated in the ground-state exciton. Thus, we can
assume a two-band model for the band structure of the system safely as long
as the thermal energy is smaller than the binding energy of the exciton. In
this case, the state of the Wannier exciton can be specified as

\begin{equation}
\left| k_{z},n\right\rangle =\sum_{l\rho }\frac{1}{\sqrt{N}}\exp
(ik_{z}r_{c})F_{n}(l),
\end{equation}
where the coefficient $1/\sqrt{N}$ is for the normalization of the state $%
\left| k_{z},n\right\rangle ,$ $k_{z}$ is the crystal momentum along the
chain direction characterizing the motion of the exciton, $n$ is the quantum
number for the internal structure of the exciton, and, in the effective mass
approximation, $r_{c}=\frac{m_{e}^{*}(l+\rho )+m_{h}^{*}\rho }{%
m_{e}^{*}+m_{h}^{*}}$ is the center of mass of the exciton. $F_{n}(l)$ is
the hydrogenic wave function with $l+\rho $ and $\rho $ being the positions
of the electron and hole, respectively. Here, $m_{e}^{*}$ and $m_{h}^{*}$
are the effective masses of the electron and hole, respectively. The
Hamiltonian for the exciton is

\begin{equation}
H_{ex}=\sum_{k_{z}n}E_{k_{z}n}c_{k_{z}n}^{\dagger }c_{k_{z}n},
\end{equation}
where $c_{k_{z}n}^{\dagger }$ and $c_{k_{z}n}$ are the creation and
destruction operators of the exciton, respectively. $E_{k_{z}n}$ is the
exciton dispersion.

The Hamiltonian of free photons is

\begin{equation}
H_{ph}=\sum_{{\bf q}^{\prime }k_{z}^{\prime }}\hbar c(q^{\prime
2}+k_{z}^{\prime 2})^{1/2}b_{{\bf q}^{\prime }k_{z}^{\prime }}^{\dagger }b_{%
{\bf q}^{\prime }k_{z}^{\prime }},
\end{equation}
where $b_{{\bf q}^{\prime }k_{z}^{\prime }}^{\dagger }$ and $b_{{\bf q}%
^{\prime }k_{z}^{\prime }}$ are the creation and destruction operators of
the photon, respectively. The wave vector ${\bf k}^{\prime }$\hspace{0.06in}%
of the photon is separated into two parts: $k_{z}^{\prime }$ is the parallel
component of ${\bf k}^{\prime }$ along the linear chain such that $k^{\prime
2}=q^{\prime 2}+k_{z}^{\prime 2}$.

In the resonance approximation\cite{10}, the interaction between the exciton
and the photon can be written in the form

\begin{equation}
H^{\prime }=\sum_{k_{z}n}\sum_{{\bf q}^{\prime }}D_{{\bf q}^{\prime
}k_{z}n}b_{k_{z}{\bf q}^{\prime }}c_{k_{z}n}^{\dagger }+{\bf h.c.,}
\end{equation}
where

\begin{equation}
D_{{\bf q}^{\prime }k_{z}n}=\frac{e}{mc}\sqrt{\frac{2\pi \hbar cN}{%
(q^{\prime 2}+k_{z}^{2})^{1/2}v}}\epsilon _{{\bf q}^{\prime }k_{z}}\chi
_{k_{z}n}
\end{equation}
with $\epsilon _{{\bf q}^{\prime }k_{z}}$ being the polarization of the
photon. In eq. (5),

\begin{equation}
\chi _{k_{z}n}=\sum_{l}F^{*}(l)\int d\tau \omega _{c}(\tau -l)\exp
(ik_{z}(\tau -\frac{m_{e}^{*}}{m_{e}^{*}+m_{h}^{*}}l))(-i\hbar \frac{%
\partial }{\partial \tau })\omega _{v}(\tau )
\end{equation}
is the effective transition dipole matrix element between the electronic
Wannier state $\omega _{c}$ in the conduction band and the Wannier hole
state $\omega _{v}$ in the valence band.

Now, we assume that at time $t=0$ the Wannier exciton is in the mode $%
k_{z},n.$ For time $t>0$,\hspace{0.06in}the state $\left| \psi
(t)\right\rangle $ for the whole system composed of the exciton and photons
can be written as

\begin{equation}
\left| \psi (t)\right\rangle =f_{0}(t)\left| k_{z},n;0\right\rangle +\sum_{%
{\bf q}^{\prime }}f_{G;{\bf q}^{\prime }k_{z}}(t)\left| G;{\bf q}^{\prime
}k_{z}\right\rangle ,
\end{equation}
where $\left| k_{z},n;0\right\rangle $ is the state with a Wannier exciton
in the mode $k_{z},n$ in the linear chain without photons, and $\left| G;%
{\bf q}^{\prime }k_{z}\right\rangle $ represents the state in which the
electron-hole pair recombines and a photon in the mode ${\bf q}^{\prime }%
{\bf ,}k_{z}$ is created.

By the method of Heitler and Ma in the resonance approximation, the
probability amplitude $f_{0}(t)$ can be expressed as\cite{10}

\begin{equation}
f_{0}(t)=\exp (-i\Omega _{k_{z}n}t-\frac{1}{2}\gamma _{_{k_{z}n}}t),
\end{equation}
where

\begin{equation}
\gamma _{_{k_{z}n}}=2\pi \sum_{{\bf q}^{\prime }}\left| D_{{\bf q}^{\prime
}k_{z}n}\right| ^{2}\delta (\omega _{{\bf q}^{\prime }k_{z}n})
\end{equation}
and

\begin{equation}
\Omega _{k_{z}n}={\cal P}\sum_{{\bf q}^{\prime }}\frac{\left| D_{{\bf q}%
^{\prime }k_{z}n}\right| ^{2}}{\omega _{{\bf q}^{\prime }k_{z}n}}
\end{equation}
with $\omega _{{\bf q}^{\prime }k_{z}n}=E_{k_{z}n}/\hbar -c\sqrt{q^{\prime
2}+k_{z}^{2}}.$ Here $\gamma _{_{k_{z}n}}$ and $\Omega _{k_{z}n}$ are,
respectively, the decay rate and frequency shift of the exciton. And ${\cal P%
}$ means the principal value of the integral.

The Wannier exciton decay rate in the optical region can be calculated
straightforwardly and is given by

\begin{equation}
\gamma _{k_{z}n}=\left\{ 
\begin{array}{l}
\frac{3\pi }{2k_{0}d}\gamma _{0}\frac{\left| {\bf \epsilon }_{{\bf q}%
^{\prime }k_{z}}\chi _{n}\right| ^{2}}{\left| \chi _{n}\right| ^{2}},\text{ 
\hspace{0.06in}}\sqrt{q^{\prime 2}+k_{z}^{2}}<k_{0} \\ 
0,\text{ \hspace{0.06in}otherwise}
\end{array}
\right. ,
\end{equation}
where $k_{0}=E_{k_{z}n}/\hbar $,

\begin{equation}
\chi _{n}=\sum_{l}F_{n}^{*}(l)\int d\tau w_{c}(\tau -l)(-i\hbar \frac{%
\partial }{\partial \tau })w_{v}(\tau ),
\end{equation}
and

\begin{equation}
\gamma _{0}=\frac{4e^{2}\hbar k_{0}}{3m^{2}c^{2}}\left| \chi _{n}\right|
^{2}.
\end{equation}

Here, $\chi _{n}^{\ast }$ represents the effective dipole matrix element for
an electron jumping from the excited Wannier state in the conduction band
back to the hole state in the valence band, and $\gamma _{0}$ is the decay
rate of an isolated atom. We see from eq.(11) that $\gamma _{_{k_{z}n}}$ is
proportional to $1/(k_{0}d)$. This is just the superradiance factor coming
from the coherent contributions of atoms within half a wavelength or so\cite
{8,18}$.$

Now let us consider the quantum wire embedded in perfectly reflecting
mirrors with cavity length $L_{c}$. If the mirror plane is parallel to the
quantum wire, it means the exciton can only couple to discrete photon modes (%
$\frac{2\pi }{L_{c}}n_{c}$, where $n_{c}$ are integers) in the perpendicular
direction, while the photon modes are still continuous in the other
direction. Following the above derivation, the decay rate can be evaluated as

\begin{equation}
\gamma _{k_{z}=0,n}=\frac{2\pi e^{2}\hbar }{m^{2}c^{2}d}%
\sum_{n_{c}=1}^{N_{c}}\frac{1}{L_{c}}\frac{\left| \epsilon _{{\bf q}^{\prime
}k_{z}}\chi _{n}\right| ^{2}}{\sqrt{k_{0}^{2}-(\frac{2\pi }{L_{c}}n_{c})^{2}}%
},
\end{equation}
where $N_{c}$ is an integer. Because of the conservation of energy, the
value of $N_{c}$ must be smaller than $L_{c}/\lambda $, where $\lambda $ is
the wavelength of the emitted photon.

As can be seen from equation (14), the exciton modes with $k_{0}<\frac{2\pi 
}{L_{c}}$ have vanishing decay rate. These exciton modes do not radiate at
all and photon trapping occurs. These {\em dark modes} also occur in a 2D
thin film. However, it is hard to examine them directly because of the
randomness of crystal momentum in a thin film. With the recent developments
of fabrication technology, it is now possible to fabricate the planar
microcavities incorporating quantum wires\cite{19}. If the thickness $L_{c}$
is equal to the wavelength of the photon emitted by bare exciton(without
external field), one can examine the dark mode directly by changing $k_{0}$
with external field.

One might argue that the singularities in eq.(14) are irrelevant in the weak
coupling region. However, there are always some leakages from the
environment in realistic systems. In this case, the summation of the
discrete modes in eq.(14) becomes the integration of the continuous modes

\begin{equation}
\gamma _{k_{z}=0,n}=\frac{e^{2}\hbar }{m^{2}c^{2}d}\int G(k_{x})\frac{\left|
\epsilon _{{\bf q}^{\prime }k_{z}}\chi _{n}\right| ^{2}}{\sqrt{%
k_{0}^{2}-k_{x}{}^{2}}}dk_{x},
\end{equation}
where the factor $G(k_{x})$ contains the informations of the leakages. For
good reflecting mirrors, we further assume $G(k_{x})$ has the form of
Lorentzian distribution and can be expressed as

\begin{equation}
G(k_{x})=N_{G}\sum_{n_{c}}\frac{1}{(k_{x}-2\pi n_{c}/L_{c})^{2}+\Delta
k_{x}^{2}},
\end{equation}
where $N_{G}$ is the normalization constant and $\Delta k_{x}$ is the line
width. Fig. 1 shows the numerical calculations of eq.(15) and (16) with $%
\Delta k_{x}$ being equal to $1\%$ of the fundamental mode $2\pi /L_{c}$. As
can be seen from the figure, the singularities smear out because of the
leakages. Therefore, as long as the corresponding band gap frequency $%
E_{k_{z}n}/\hbar $ is much larger than the decay rate at these peak values,
the perturbation theory still works well. One also notes the peak value
decreases with the increasing of cavity length. Therefore, the decay rate
should approach to the free space limit in eq.(11) as the cavity length
becomes infinity. In the work of Ref.[19], C. Constantin {\em et al}.
investigated the transition from nonresonant mode to resonant coupling
between quantum confined one-dimensional carriers and two-dimensional photon
states in a planar Bragg microcavity with a size of one wavelength
incorporating strained In$_{0.15}$Ga$_{0.85}$/GaAs V-groove quantum wires.
They found that when the excitonic transition energy is resonant with the
cavity mode, the emission rate into this mode is significantly enhanced.
This significant feature is just the enhancement in Fig.1 and can be easily
explained by the present model.

To understand the emission rate increasing to resonance thoroughly, we now
consider a Wannier exciton in a quantum ring embedded in perfectly
reflecting mirrors with cavity length $L_{c}$. The circular ring is joined
by the $N_{r}$ lattice points with radius $\rho \sim N_{r}d/2\pi $, where $d$
is the lattice spacing and the number of the lattice points is $N_{r}$.
Following the above derivation, the decay rate of the quantum ring exciton
can be expressed as

\begin{equation}
\gamma _{\nu }=\sum_{n_{c}}\frac{e^{2}\hbar }{m^{2}c^{2}L_{c}}\frac{\rho }{d}%
\left| H_{\nu }^{(1)}(\sqrt{(2\pi /\lambda )^{2}-(2\pi n_{c}/L_{c})^{2}}\rho
)\right| ^{2}\left| {\bf \epsilon }_{{\bf q}^{\prime }k_{z}^{\prime }}\cdot 
{\bf \chi }_{\nu }\right| ^{2},
\end{equation}
where $\nu $ is the exciton wave number in the circular direction and $%
H_{\nu }^{(1)}$ is the Hankel function. As can be seen from eq. (17), the
decay rate of a quantum ring exciton also shows the enhanced peaks as the
cavity length $L_{c}$ is equal to the multiple wavelength of the emitted
photon. However, if one considers a Wannier exciton in a quantum dot
embedded in the microcavity, the decay rate :

\begin{equation}
\gamma \propto \sum_{n_{c}}\frac{e^{2}\hbar }{m^{2}c^{2}L_{c}}\theta ((2\pi
/\lambda )^{2}-(2\pi n_{c}/L_{c})^{2})\left| {\bf \epsilon }_{{\bf q}%
^{\prime }k_{z}^{\prime }}\cdot {\bf \chi }\right| ^{2},
\end{equation}
where $\theta $ is the step function, shows no peak, instead, only the
plateau appears with the increasing of the cavity length. This is because
the angular momentum (translational momentum) of the exciton in a quantum
ring (wire) is conserved in circular (chain) direction, while the crystal
symmetry is totally broken in a quantum dot. Due to the modification of the
density of states of the photon in the microcavity, the decay rate of the
exciton shows peaks in a quasi-one dimensional system but {\em plateaus} in
a quasi-zero dimensional system. One should notice that such kind of peak
maybe a useful feature to realize the Aharonov-Bohm effect for an exciton in
a quantum ring. Recently, R\"{o}mer and Raikh\cite{22} studied theoretically
the exciton absorption on a ring threaded by a magnetic flux. In order to
see the AB oscillations, they suggested to measure the luminescence. In this
case, however, the excitonic AB oscillations is very small and hard to be
measured. Therefore, if one can incorporate the quantum ring with planar
microcavities, the AB oscillations will be enhanced at these peaks.

A few remarks about frequency shift $\Omega _{k_{z}n}$ in a quantum wire can
be mentioned here. In perfect microcavities, the frequency shift can be
expressed as

\begin{equation}
\Omega _{k_{z}=0,n}=\frac{e^{2}\hbar }{m^{2}c^{2}d}\sum_{n_{c}=1}^{N_{c}}%
\frac{1}{L_{c}}\int \frac{\left| \epsilon _{{\bf q}^{\prime }k_{z}}\chi
_{n}\right| ^{2}}{(k_{0}-\sqrt{k_{x}{}^{2}+(\frac{2\pi }{L_{c}}n_{c})^{2}})%
\sqrt{k_{x}{}^{2}+(\frac{2\pi }{L_{c}}n_{c})^{2}}}dk_{x}.
\end{equation}
As pointed out in Ref. [21], the frequency shift suffers from divergences
and has to be removed by renormalization. Following the renormalization
procedure proposed by Lee {\em et al}.\cite{24}, one can, in principle,
obtain the frequency shift in eq. (19). In the free space limit, the
frequency shift can be approximated as

\begin{equation}
\Omega _{k_{z}=0,n}^{ren}\sim -\gamma _{\sin gle}(\frac{1}{k_{0}d}),
\end{equation}
where $\gamma _{\sin gle}$ is roughly equal to the decay rate of an isolated
atom\cite{24}. Similar to the decay rate in eq.(11), the frequency shift is
also superradiatively enhanced by the coherent effect. The numerical
calculations of eq.(19) are shown in Fig. 2. As can be seen from the figure,
the frequency shift has discontinuities when the cavity length is equal to
the multiple wavelength of the emitted photon. This is because whenever the
cavity length exceeds some multiple wavelength, it opens up another decay
channel abruptly. These discontinuities should also smear out because of the
leakages from the environment. One can also note that as the cavity length
increases, the sign of the frequency shift changes from positive to negative
and approaches the free space limit. This kind of crossing also occurs in
the quantum well systems\cite{23}, and can be realized by the competition
between the negative and positive values of the integration in eq.(19).

For usual semiconductors, the enhanced factor in eq. (20) is about $10^{3}$
for Wannier excitons in the optical range. However, due to the extreme
smallness of $\gamma _{\sin gle}$ itself, observation of $\Omega _{k_{z}\sim
0,n}^{ren}$ is not expected to be easy. The discontinuous behavior at
certain cavity length $L_{c}$ may be a useful feature to observe this
quantity. Besides, we suggest to investigate the semiconductor materials
which have a larger exciton oscillator strength and thus a larger frequency
shift. As for the magnitude of this shift, if the decay rate of the exciton
is in the order of ps$^{-1}$, the radiative shift is about $10^{-1}$meV. One
can also vary the number ($N_{0}$) of the wires. Due to the coherent
effects, the measured frequency shift would be $N_{0}\Omega _{k_{z}\sim
0,n}^{ren}$ if the wires are placed within the coherent length. However, one
should also note the inhomogeneous broadening caused by the fluctuations in
the wire sizes which cannot be kept absolutely constant, leading to the
small band gap fluctuation of quantum wires. The inhomogeneous broadening
may be taken into account by assuming that the wires have a size
distribution given by $f(E_{k_{z}})$ around a mean value $\overline{E_{k_{z}}%
}.$ The average frequency shift is then $\overline{\Omega _{k_{z}}^{ren}}%
=\int f(E_{k_{z}})\left. \Omega _{k_{z}\sim 0,n}^{ren}\right|
_{E_{k_{z}}}dE_{k_{z}}$.

In summary, we have calculated the decay rate of the Wannier exciton in a
quantum wire. When the quantum wire is incorporated into planar
microcavities, it becomes possible to examine the dark modes of the exciton.
Besides, the decay rate is greatly enhanced as the cavity length $L_{c}$ is
equal to the multiple wavelength of the emitted photon. The singularities in
the decay rate smear out as a result of leakages from the environment.
Furthermore, we have also calculated the renormalized frequency shift of the
exciton. Similar to its decay rate counterpart, the frequency shift shows
discontinuity at resonant modes. The distinguishing features are pointed out
and may be observable in a suitably designed experiment.

This work is supported partially by the National Science Council, Taiwan
under the grant number NSC 90-2112-M-009-018.

\newpage

\subsection{FIGURE CAPTION}

Fig. 1 Decay Rate of the superradiant exciton in a quantum wire embedded in
cavities with leakages$.$ The vertical and horizontal units are $\frac{2\pi
e^{2}\hbar \left| {\bf \epsilon }_{{\bf q}^{\prime }k_{z}}\chi _{n}\right|
^{2}}{m^{2}c^{2}k_{0}d}$ and $\lambda ,$ respectively.

Fig. 2 Renormalized frequency shift of the superradiant exciton in a quantum
wire as a function of the cavity length $L_{c}.$ The vertical and horizontal
units are $\frac{e^{2}\hbar \left| {\bf \epsilon }_{{\bf q}^{\prime
}k_{z}}\chi _{n}\right| ^{2}}{m^{2}c^{2}k_{0}d}$ and $\lambda ,$
respectively.

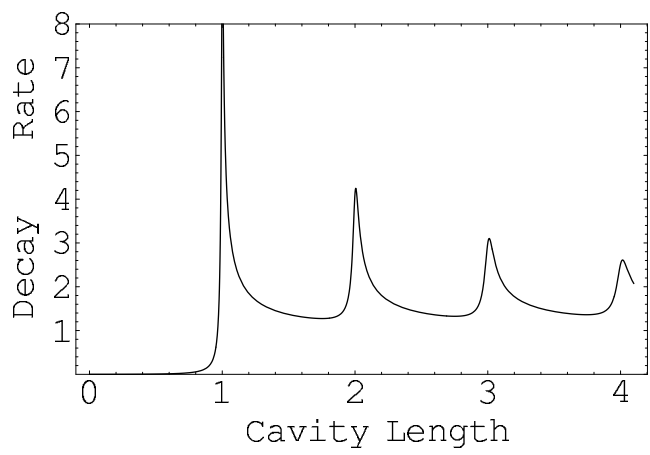
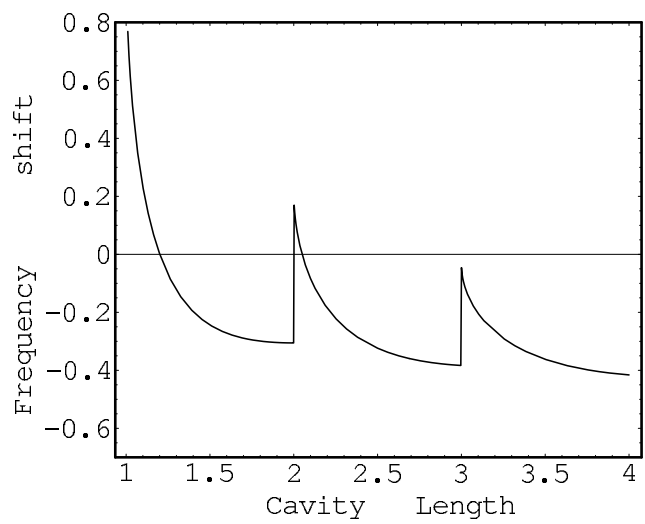
\end{document}